\documentclass[12pt]{article} 
\usepackage{amssymb} 

\usepackage[tablesfirst,notablist,nomarkers]{endfloat} 

\usepackage{ol2}
\usepackage[draft]{hyperref}
\usepackage{amsmath}

\begin{document}


\title{Two-dimensional subwavelength plasmonic lattice solitons }

\author{F. Ye,$^{1,5}$ D. Mihalache,$^2$ B. Hu,$^{3,5}$ and N. C. Panoiu$^{4}$}

\address{
$^1$Department of Physics, Shanghai Jiao Tong University, Shanghai
200240, China
\\
$^2$"Horia Hulubei"National Institute for Physics and Nuclear
Engineering, Department of Theoretical Physics, 407 Atomistilor,
Magurele-Bucharest, 077125, Romania \\
$^3$Department of Physics, University of Houston, Houston, Texas 77204-5005, USA \\
$^4$Department of Electronic and Electrical Engineering, University College London, Torrington Place, London WC1E 7JE, UK \\
$^5$Centre for Nonlinear Studies, and The Beijing-Hong
Kong-Singapore Joint Centre for Nonlinear and Complex Systems (Hong
Kong), Hong Kong Baptist University, Kowloon Tong, Hong Kong, China

}

\begin{abstract} We present a theoretical study of plasmonic lattice solitons (PLSs) formed in two-dimensional (2D) arrays
of metallic nanowires embedded into a nonlinear medium with Kerr nonlinearity. We analyze two classes of 2D PLSs families,
namely, fundamental and vortical PLSs in both focusing and defocusing media. Their existence, stability, and subwavelength
spatial confinement are studied in detail.\end{abstract}

\noindent When the size of photonic systems is reduced to subwavelength scale, the light confinement and propagation are
severely limited by diffraction. Overcoming this major challenge of confining and manipulating optical energy at the
nanoscale has attracted a rapidly growing interest in nanophotonics \cite{bde03n}. To this end, one promising approach
employs surface plasmon polaritons (SPPs) of metallo-dielectric nanostructures\cite{zsm05pr}. Although most of the
experimental and theoretical studies of SPPs have focused on their linear optical features, there is a growing interest in
the optical properties of SPPs in the nonlinear regime\cite{fo07ol,lbg07prl,kivshar08oe,fye10prl,skryabin10ol}. This is
primarily because the strong enhancement of the optical field induced by the excitation of SPPs can significantly boost
nonlinear optical effects, as well as the promising applications of nonlinear SPPs in the all-optical control at nanoscale.
A new approach to achieve subwavelength confinement and control of the optical field, employing one-dimensional (1D) arrays
of parallel metallic nanowires embedded in an optical medium with Kerr nonlinearity, was recently proposed\cite{fye10prl}.
Under certain conditions the optical nonlinearity induced by SPPs field exactly balances the discrete diffraction in the
plasmonic array, and as such collective excitation of 1D {\it plasmonic lattice solitons} (PLSs), are formed in the array.
Because the radius $a$ of the nanowires and their separation distance $d$, are much smaller than the operating wavelength
$\lambda$, the spatial width of the PLSs can be significantly smaller than $\lambda$.

\begin{figure}[htb]
\centerline{\includegraphics[width=8cm]{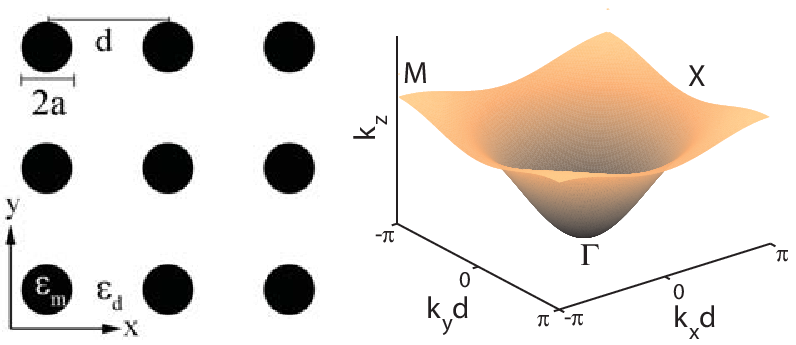}} \caption{(Color online) (a) Schematics of a 2D square array of nanowires
with radius $a$ and separation distance $d$. (b) The band predicted by the coupled mode theory.}\label{geom}
\end{figure}

In this Letter, we study the nonlinear optical modes in 2D arrays of metallic nanowires which are embedded into a Kerr
medium [Fig.~\ref{geom}(a)]. We found that 2D PLSs of subwavelength spatial extent in the transverse section could exist in
this geometry. In addition to the fundamental PLSs, 2D plasmonic arrays support vortical solitons, i.e., solitons with an
internal topological phase change of $2\pi$ along a closed contour around soliton's phase singularity. The vortical PLSs can
also be of subwavelength extent. Fundamental and vortical PLSs are found to exist in both focusing and defocusing media.

Our analysis is an extension of the coupled mode theory developed in Ref.\cite{fye10prl} for 1D coupled metallic nanowires.
Briefly, the total electric and magnetic fields are expanded as a superposition of the modes of a single nanowire (assumed
to have only the fundamental TM mode), and using the conjugated form of the Lorentz reciprocity theorem one obtains
\begin{align}\label{model}
i&\frac{\displaystyle d\phi_{m,n}}{\displaystyle dz}+\kappa(\phi_{m,n+1}+\phi_{m,n-1}+\phi_{m+1,n}+\phi_{m-1,n})+ \notag \\
&\mu(\phi_{m+1,n+1}+\phi_{m+1,n-1}+\phi_{m-1,n+1}+\phi_{m-1,n-1})+\notag \\&\gamma\vert\phi_{m,n}\vert^2\phi_{m,n}=0,
\end{align}
where $z$ is the longitudinal coordinate, $\phi_{m,n}$ is the normalized mode amplitude in nanowire $(m,n)$, $\kappa$ and
$\mu$ are the coupling coefficients between neighboring and next-neighboring nanowires, respectively, and $\gamma$ is the
normalized nonlinearity coefficient. Note that $\gamma>0$ ($\gamma<0$) for a medium with self-focusing (self-defocusing)
Kerr response. We emphasize that, unlike in the case of conventional dielectric waveguides, for our plasmonic array,
$\kappa<0$, $\mu<0$, so that the linear dispersion, $k_z=2\kappa[cos(k_xd)+cos(k_yd)]+4\mu cos(k_xd)cos(k_yd)$, forms a
concave surface with anomalous diffraction occurring at the $\Gamma$ point ($k_x=k_y=0$), while normal diffraction at the
$M$ points ($k_x=k_y=\pm\pi/d$) [Fig.~\ref{geom}(b)].

The soliton solution to Eq. (\ref{model}) are sought in the form $\phi_{m,n}(z)=u_{m,n}\exp(i\beta\kappa z)$, where the
amplitudes $u_{m,n}$ are independent of $z$ and $\beta$ is the soliton wave number (normalized by $\kappa$). The soliton is
characterized by its power $P=\sum_{m,n}|\phi_{m,n}|^2$. This \textit{ansatz} is inserted into Eq. (\ref{model}) and the
resulting system of equations is solved numerically by a standard relaxation method. The corresponding amplitudes are then
used to reconstruct the fields of the PLSs. To quantify the transverse size of solitons we use the effective radius,
$R=(\iint_{-\infty}^{\infty} |\textbf{E}|^2[{(x-x_0)}^2+{(y-y_0)}^2]dxdy /\iint_{-\infty}^{\infty}|\textbf{E}|^2
dxdy)^{1/2}$, where $(x_0,y_0)=\iint_{-\infty}^{\infty}(x,y)|\textbf{E}|^2dxdy/\iint_{-\infty}^{\infty} |\mathbf{E}|^2dxdy$
is the mean center position. In the analysis below, we assumed that the nanowires are made of Ag and used the Drude model,
$\epsilon_m(\omega)=1-\frac{\omega_{p}^{2}}{\omega(\omega+i\nu)}$, for the permittivity of the metal. For Ag,
$\omega_p=13.7\times 10^{15}~\mathrm{rad/s}$ and $\nu=2.7\times 10^{13}~\mathrm{rad/s}$. For simplicity, we assume that
metal is lossless.

\begin{figure}[htb]
\centerline{\includegraphics[width=8cm]{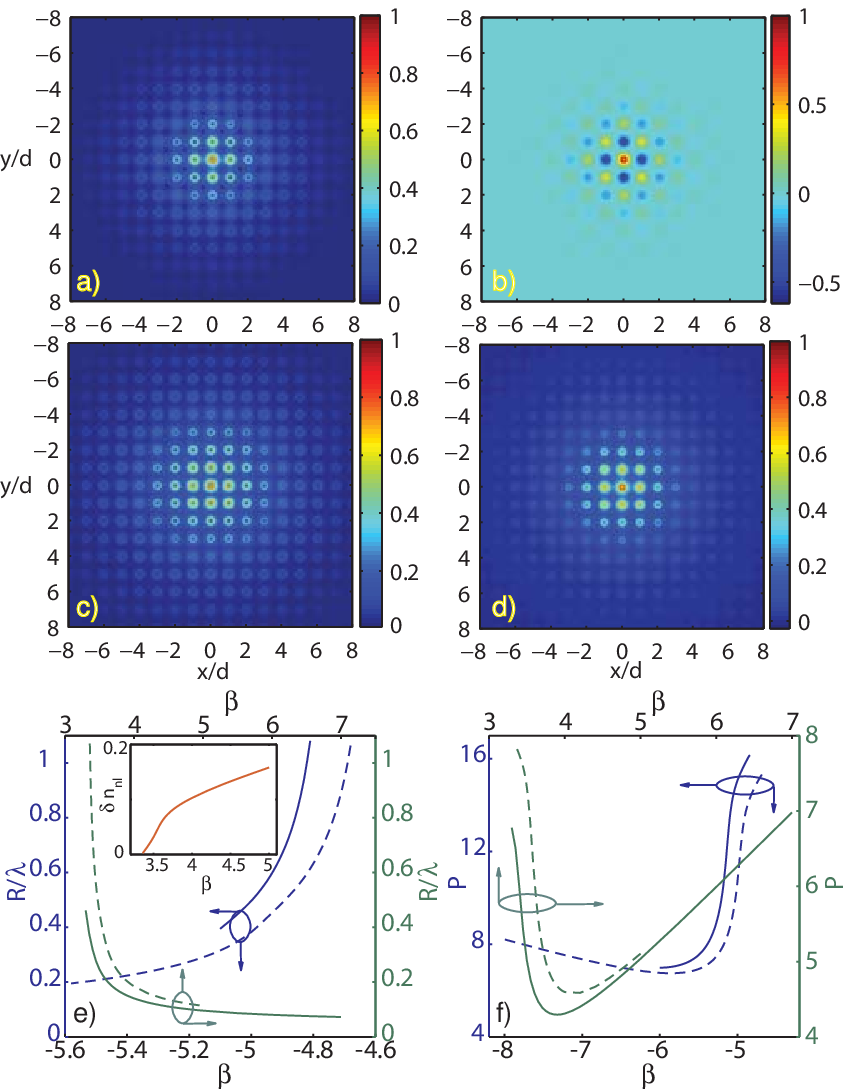}} \caption{(Color
online) The normalized electric field amplitude (a,c) and
longitudinal component (b,d) in focusing (top panels) and defocusing
(middle panels) media. $\lambda=1550~\mathrm{nm}$,
$a=40~\mathrm{nm}$, $d=8a$. (a,b) $\beta=3.58$, $\delta
n_{\mathrm{nl}}=0.05$. (c,d) $\beta=-4.89$, $\delta
n_{\mathrm{nl}}=-0.05$. Radius $R$ (e) and power $P$ (f) vs $\beta$.
The solid and dashed lines stand for $d=7a$ and $d=8a$,
respectively. In inset, $\delta n_{\mathrm{nl}}$ vs $\beta$ for
$d=8a$, in focusing medium.}\label{fdtd}
\end{figure}

We first search for fundamental solitons. As in the case of the 1D
PLSs \cite{fye10prl}, we also find two types of 2D PLSs, i.e.,
staggered and unstaggered solitons. In the case of unstaggered
(staggered) PLSs the phase difference of the mode amplitude in
adjacent nanowires is equal to zero ($\pi$). The spatial profile of
the amplitude and longitudinal field of unstaggered (staggered) PLSs
corresponding to a nonlinear change of $\delta
n_{\mathrm{nl}}=-0.05$ ($\delta n_{\mathrm{nl}}=0.05$) are shown in
Fig.~\ref{fdtd}(a) and (b) [Fig.~\ref{fdtd}(c) and (d)].
Importantly, due to the inverted linear dispersion relation,
staggered (unstaggered) solitons are formed in self-focusing
(self-defocusing) media, which is opposite to the case of dielectric
waveguide arrays. Note also that the solitons have subwavelength
size, the radius being $R=0.35\lambda$ [Fig.~\ref{fdtd}(a,b)] and
$R=0.50\lambda$ [Fig.~\ref{fdtd}(c,d)]. The dependence of soliton
radius on $\beta$ is shown in Fig.~\ref{fdtd}(e). The plot shows
that both staggered and unstaggered PLSs shrink monotonically as
their wave number moves away from the band edges. This is because
staggered (unstaggered) PLSs originate from the (extended) Bloch
modes at the point $M$ ($\Gamma$), and thus when their wave number
is close to the band edge, solitons have a large width. A larger
change of nonlinear refractive index tunes the soliton mode deeper
into the gap [see inset of Fig.~\ref{fdtd}(e)] and thus PLSs are
increasingly smaller than $\lambda$. Importantly, our calculations
show that the power carried by such PLSs cannot be smaller than a
threshold value, as can be seen from Fig.~\ref{fdtd}(f). The
nonmonotonic relationship $P=P(\beta)$ has an important implication
on soliton stability, as will be shown below.

\begin{figure}[t]
\centerline{\includegraphics[width=8cm]{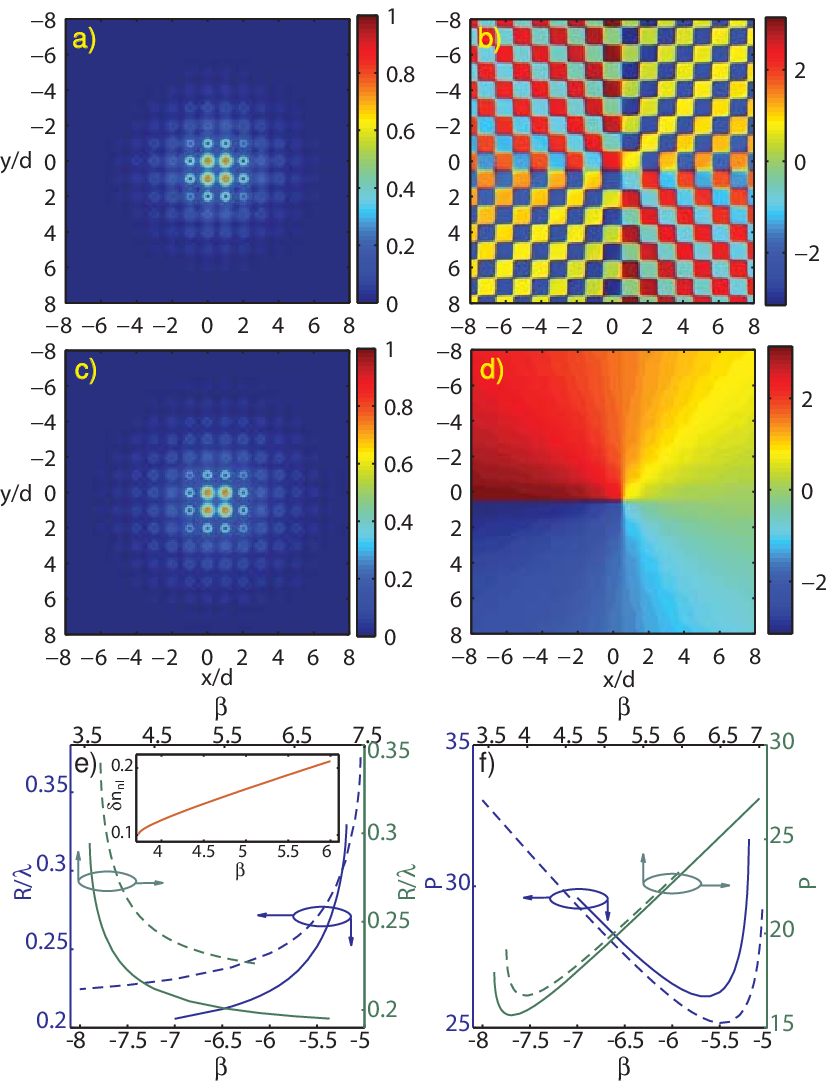}} \caption{(Color
online) The normalized electric field amplitude (a,c) and
longitudinal component (b,d) in focusing (top panels) and defocusing
(middle panels) nonlinear media. (a,b) $\beta=3.82$, $\delta
n_{\mathrm{nl}}=0.11$. (c,d) $\beta=-5.2$, $\delta
n_{\mathrm{nl}}=-0.18$. Radius $R$ (e) and power $P$ (f) vs
normalized eigenvalue $\beta$. The solid and dashed lines stand for
$d=7a$ and $d=8a$, respectively. In inset, $\delta n_{\mathrm{nl}}$
vs $\beta$ for $d=8a$, in focusing medium.}\label{vor}
\end{figure}

In addition to the fundamental PLSs, 2D plasmonic arrays also support a new type of PLSs, which has no counterpart in 1D
systems, i.e., vortex PLSs; these belong to the class of \textit{discrete vortices} \cite{Boris}. Here we focus on off-site
vortices, i.e., compact vortex states whose singularity is located between lattice sites. Similarly to fundamental PLSs,
vortex PLSs exist both in self-focusing [Fig.~\ref{vor}(a,b)] and self-defocusing media [Fig.~\ref{vor}(c,d)]. As can be
easily seen from the phase profiles, the vortex has a topological charge equal to 1 (2D PLSs with topological charge 2 also
exist, although not shown in here). Interestingly, the phase profile for PLSs in focusing media features a "staggered"
pattern at the sites far away from the soliton center, which is a typical feature of "gap" vortices \cite{gapvor04prl}.
However, one should mention that the staggered vortex PLS resides at the semi-infinite gap (above the band) and thus is not
a "gap" vortex. This is again a consequence of the inverted linear dispersions. The properties of vortex PLSs are summarized
in Fig.~\ref{vor}(e,f). Several differences from fundamental PLSs are observed. First, unlike the fundamental PLSs where
their value of $\beta$ can be very close to the band edges, vortex soliton can only exist in the gaps at a finite distance
from the band edges. This is due to the fact that the vortex mode is not a Bloch mode of the linear plasmonic array.
Secondly, as vortex PLSs possess four intensity maxima, vortex solitons always have a larger width as compared to that of
fundamental ones (under the same change of nonlinear refractive index). Therefore, forming a subwavelength vortex generally
requires stronger nonlinearity. Nevertheless, we find that subwavelength vortex PLSs can be achieved under experimentally
accessible conditions. For example, the vortex presented in Fig.~\ref{vor}(a,b) has radius $R=0.3\lambda$ requiring a
nonlinear change of refractive index $\delta n_{\mathrm{nl}}=0.11$. Note that an index change of $\sim$0.14 was reported in
Ref. \cite{bs03apl}. Further, the PLSs size can be significantly reduced if the operating wavelength is scaled down, thus
the requirement for a strong nonlinear change of the refractive index can be relaxed. Finally, we mention that as in the
case of fundamental solitons, vortex PLSs also feature power thresholds for their formation.

\begin{figure}[t]
\centerline{\includegraphics[width=8cm]{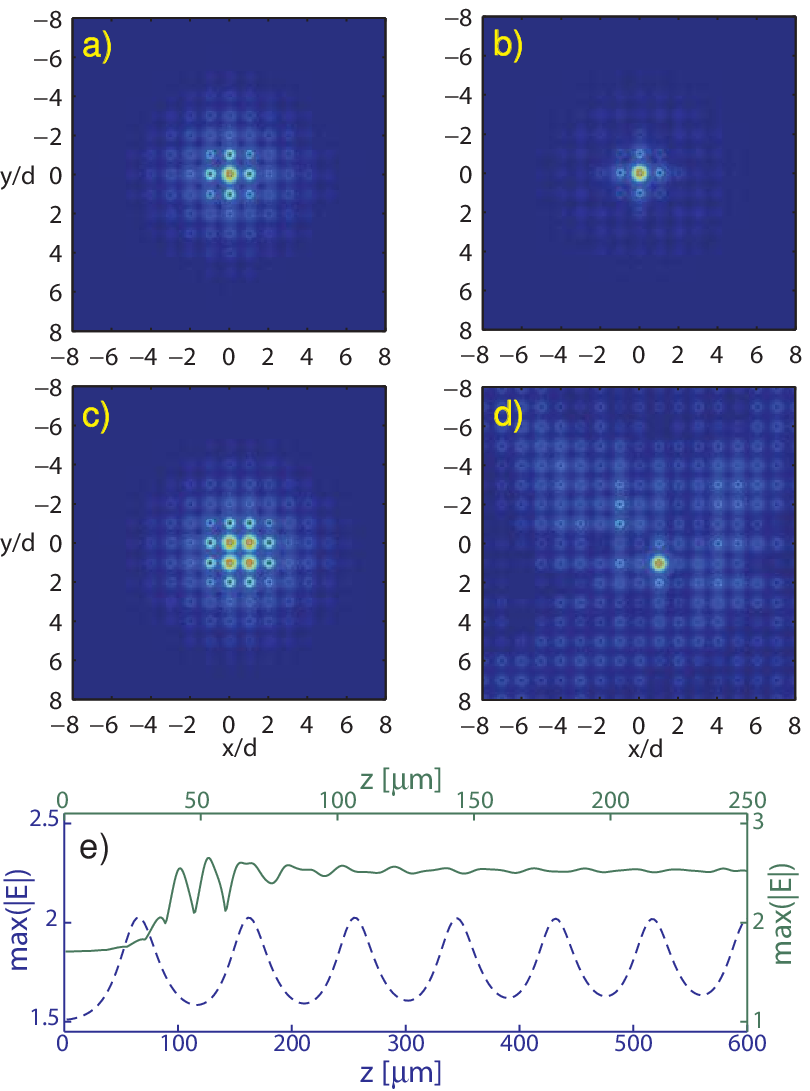}} \caption{(Color online) Unstable propagation of the fundamental soliton
at $\beta=3.8$ (top panels) and vortex at $\beta=3.82$ (middle panels), showing the intensity profile at $z=0$ (a,c),
$z=60~\mu \mathrm{m}$ (b), and $z=90~\mu \mathrm{m}$ (d). (e) Variation of maximum amplitude vs propagation distance for
fundamental (dashed) and vortex (solid) solitons.}\label{prop}
\end{figure}

A relevant issue associated with the 2D PLSs is their stability. To
address this issue, we integrate Eq. (\ref{model}) by using the
Runge-Kutta method with the input condition being the stationary
soliton solution plus random noise. Our extensive simulations show
that the fundamental PLSs obey the Vakhitov-Kolokolov stability
criterion namely, staggered (unstaggered) PLSs are stable when
$dP/d\beta>0$ ($dP/d\beta<0$) and are unstable otherwise. In the
stability region, soliton maintains its original shape over a
sufficiently long distance while the initial added noise is rapidly
eliminated. In contrast, unstable solitons tend to relax to stable
ones with the same power and large oscillations are observed during
propagation [Fig.~\ref{prop}(a,b,e)]. The vortex can be also stable,
although its stability region is more limited. An unstable
propagation of vortex PLS is shown in Fig.~\ref{prop}(c,d). The
vortex quickly loses its screwed phase structure and decays into a
fundamental PLS. Note that the stability domain of vortex PLSs
increases at a shorter operating wavelength.

In conclusion, we presented the first study of 2D plamonic lattice
solitons in arrays of metallic nanowires embedded into a host medium
with Kerr nonlinearity. Stable subwavelengh fundamental and vortex
PLSs are found to exist in both focusing and defocusing media. The
work of N. C. P. was supported by the EPSRC.

\end{document}